\documentclass[prb,preprint,showpacs,preprintnumbers,amsmath,amssymb]{revtex4}

\usepackage{graphicx}

\usepackage{dcolumn}

\usepackage{bm}

\newcommand{\gma}{$\rm{Ga_{1-x}Mn_{x}As}$}

\newcommand{\rxx}{$R_{\rm{xx}}$}
\newcommand{\rxy}{$R_{\rm{xy}}$}

\begin{document}
\title{The Influence of Magnetic Domain Walls on Longitudinal and Transverse Magnetoresistance in Tensile Strained (Ga,Mn)As Epilayers}
\author{G.\ Xiang}
\author{N.\ Samarth}\email{nsamarth@psu.edu}
\affiliation{Physics Department $\&$ Materials Research Institute, Penn State University, University Park PA 16802}

\begin{abstract}
We present a theoretical analysis of recent experimental measurements of magnetoresistance in \gma~ epilayers with perpendicular magnetic anisotropy. The model reproduces the field-antisymmetric anomalies observed in the longitudinal magnetoresistance in the planar geometry (magnetic field in the epilayer plane and parallel to the current density), as well as the unusual shape of the accompanying transverse magnetoresistance. The magnetoresistance characteristics are attributed to circulating currents created by the presence of magnetic domain walls. 

\end{abstract}
\maketitle

\section {Introduction}
Contemporary developments in spintronics \cite{wolf_science_2001}
have generated much interest in the interplay between electrical transport and magnetic domain walls (DWs) in ferromagnetic metals \cite{kent_review,ono_science_1999,allwood_science_2002,cheng_prl_2005} 
and in ferromagnetic 
semiconductors.\cite{tang_prl_2003,tang_nature_2004,holleitner_apl_2004,yamanouchi_nature} It is of particular relevance in this context to study the relationship between spin transport and DWs in the ``canonical" ferromagnetic semiconductor \gma\cite{spinbook,samarth_SSP,nature_mat} because this material is a model system for proof-of-concept semiconductor spintronic devices.\cite{awschalom_naturephys,xiang_conmatt0607580} Indeed, recent experimental studies of \gma~ devices have demonstrated that the presence of DWs directly influences measurements of the longitudinal magnetoresistance (\rxx$(H)$) due to contributions from the transverse magnetoresistance (\rxy$(H)$): 
in samples with in-plane magnetic anisotropy, this arises because of the giant planar Hall effect,\cite{tang_prl_2003,tang_nature_2004}while in samples with perpendicular magnetic anisotropy, the admixture is created by the anomalous Hall effect.\cite{xiang_prb_2005} In the latter case, \rxx$(H)$ shows remarkable field-antisymmetric anomalies in the {\it planar geometry} when the external magnetic field ($\vec{H}$) is applied parallel to the current density ($\vec{j}$) and perpendicular to the magnetic easy axis ($\hat{z}$). This is similar to observations of a field-antisymmetric magnetoresistance in metallic ferromagnetic multilayers with perpendicular anisotropy, but we note that in the metallic case the external field was applied along the {\it easy} axis.\cite{cheng_prl_2005} Theoretical modeling has shown how circulating currents in the vicinity of a DW can result in an admixture of \rxy$(H)$ in the measurement of \rxx$(H)$ for $\vec{H}$ applied along the easy axis of a ferromagnetic thin film.\cite{tang_prb_2004, cheng_prl_2005}. In this paper, we extend these calculations to a different experimental geometry where $\vec{H}$ is applied along a {\it hard axis} of a ferromagnetic thin film with perpendicular magnetic anisotropy. The magnetization reversal process in this case involves a 3 dimensional process rather than a 1 or 2 dimensional one as in the calculations published earlier. Our theoretical analysis is aimed at explaining the unusual hard axis magnetoresistance observed in tensile-strained \gma~ epilayers with perpendicular magnetic anisotropy.\cite{xiang_prb_2005} 

When we measure \rxx$(H)$ and \rxy$(H)$ in tensile-strained \gma~ in the planar geometry ($\vec{H} || \vec{j} \bot \hat{z}$), we observe the following characteristics: 
\begin{enumerate}
\item  The background longitudinal magnetoresistance is symmetric with respect to the direction of the magnetic field $\vec{H}$ but there are resistance ``spikes" with an antisymmetric deviation $\Delta R(H) = -\Delta R(-H)$. These anomalies arise when the magnetization reverses.
\item The transverse magnetoresistance shows a hysteresis loop with an unusual shape, with \rxy$(H)=0$ at the field where \rxx$(H)$ shows a maximum or minimum spike. 
\end{enumerate}
The model presented in this paper shows that our observations in  tensile-strained \gma~ can be explained by the same concepts used to understand the magnetoresistance in unstrained \gma~ with planar anisotropy \cite{tang_prl_2003,tang_nature_2004,tang_prb_2004} and in metallic multilayers with perpendicular magnetic anisotropy:\cite{cheng_prl_2005} circulating currents near a DW located between the voltage probes produce Hall effect contributions to \rxx. This effect manifests itself particularly during magnetization reversal via DW nucleation and propagation.

In our model, we treat the device as a rectangular Hall bar of width {\it w} and thickness {\it t} (Fig. 1). The film is in the {\it xy} plane and the length is along the {\it x} axis. We begin with the assumption that a DW  is positioned at $x=0$ in the {\it yz} plane, separating the thin film into two domains ($i=1$ and $i=2$ to the left and right of the DW, respectively) with opposite magnetization. In each domain, the electric field and the current density are related by ${\vec E}_i  = \rho _i {\vec j}_i$, where the resistivity tensor is given by:
\begin{eqnarray}
\rho_{1}= \left( \begin{array}{cc}
              \rho& -\rho_{H}\\
                 \rho_{H}&\rho
               \end{array} \right) \\
\rho_{2}= \left( \begin{array}{cc}
              \rho& \rho_{H}\\
                 -\rho_{H}&\rho
               \end{array} \right) 
\end{eqnarray}       

In the above equations, $\rho$ and $\rho_H$ are the diagonal and off-diagonal components of the resistivity tensor; note that the latter change sign between the two domains because of the anomalous Hall effect. Equations (1) and (2) may then be written as: 
\begin{eqnarray}
\left( \begin{array}{c}
              \frac{-\partial V_{i}}{\partial x}\\
               \frac{-\partial V_{i}}{\partial y}  
               \end{array} \right)               = \left( \begin{array}{cc}
                                                 \rho& \rho_{H}{\mathop{\rm sgn}}(x)\\
                                                    -\rho_{H}{\mathop{\rm sgn}}(x)&\rho
                                                    \end{array} \right)  \times \left( \begin{array}{c}
                                                                                     j_{xi}\\
                                                                                      j_{yi}
                                                                                        \end{array} \right) \end{eqnarray}

Assuming no static charges accumulate in the \gma~ sample, the electric potential and current density satisfy the following boundary conditions :
\begin{eqnarray}
	\nabla ^2 V_i  = 0, \\ 
 j_{xi} ( \pm \infty ,y) = j_0, \\ 
 j_{yi} ( \pm \infty ,y) = 0, \\ 
 j_{yi} (0,0) = 0, \\ 
 j_{yi} (0,w) = 0.
\end{eqnarray}

Also, the continuities of the electric potential and {\it x} component of current at the interface require: 
\begin{eqnarray}
	j_{x1} (0,y) = j_{x2} (0,y), \\ 
 V_1 (0,y) = V_2 (0,y).
\end{eqnarray}
  
Using the above boundary conditions, we solve for the electric potential and the current density in the limit $\beta = \frac{\rho_{H}}{\rho} << 1$ to obtain:\cite{bate_jap_1961, partin_jap_1973}

\begin{eqnarray}
V_1 (x,y)&=&V_{10}  - j_0 \rho (x + \beta y) - \rho \sum\limits_{n = 1}^\infty  {A_n \exp (\frac{{n\pi x}}{w})} [\cos (\frac{{n\pi y}}{w}) + \sin (\frac{{n\pi y}}{w})]\\ 
V_2 (x,y)&=&V_{20}  - j_0 \rho (x - \beta y) + \rho \sum\limits_{n = 1}^\infty  {A_n \exp (-\frac{{ n\pi x}}{w})} [\cos (\frac{{n\pi y}}{w}) + \sin (\frac{{n\pi y}}{w})] \\ 
j_{x1}&=&j_0  + \frac{\pi }{w}\sum\limits_{n = 1}^\infty  {nA_n \exp (\frac{{n\pi x}}{w})} \cos (\frac{{n\pi y}}{w}) \\ 
j_{y1}&=& - \frac{\pi }{w}\sum\limits_{n = 1}^\infty  {nA_n \exp (\frac{{n\pi x}}{w})} \sin (\frac{{n\pi y}}{w})\\ 
j_{x2}&=&j_0  + \frac{\pi }{w}\sum\limits_{n = 1}^\infty  {nA_n \exp ( - \frac{{n\pi x}}{w})} \cos (\frac{{n\pi y}}{w})\\ 
j_{y2}&=&+ \frac{\pi }{w}\sum\limits_{n = 1}^\infty  {nA_n \exp ( - \frac{{n\pi x}}{w})} \sin (\frac{{n\pi y}}{w})
\end{eqnarray}

where 
\begin{eqnarray}
 A_1  &=& (j_0 w/\pi )(4\beta /\pi )[1 - 0.205(4\beta /\pi )^2  + ...], \\ 
 A_2  &=& (j_0 w/4\pi )(4\beta /\pi )[(4\beta /\pi ) - 0.412(4\beta /\pi )^3  + ...], \\ 
 A_3  &=& (j_0 w/9\pi )(4\beta /\pi )[1 + 0.297(4\beta /\pi )^2  + ...], \\ 
 A_4  &=& (j_0 w/16\pi )(4\beta /\pi )[\frac{4}{3}(4\beta /\pi ) - 0.397(4\beta /\pi )^3  + ...], \\ 
 V_{20}  - V_{10}  &=&  - j_0 w\beta \rho \{ 1 + (4/\pi ^2 )(4\beta /\pi ) \times [1.052 - 0.181(4\beta _{} /\pi )^2  + ...]\} . 
\end{eqnarray}

We note that the transverse field $E_y$ due to the Hall effect changes sign from $-\infty$ to $+\infty$. By symmetry, $E_y$ vanishes in the vicinity of $x=0$ where the domain wall is located. The Lorentz force is then not balanced due to the lack of an electric force e$E_y$, and the carriers are deflected towards one side of the sample, causing a nonuniform circulating current around the DW at $x=0$. 

\section {Simulation of the Hall Resistance}

Using the above model, we quantitatively calculate the Hall voltage at $x$ when the domain wall is located at $x=0$.
\begin{eqnarray}
	V_H (x) &=& V_i (x,0) - V_i (x,w).
\end{eqnarray}
To first order in $\beta$, the Hall voltage is 

\begin{eqnarray}
V_H (x) &=& (\rho _H j_0 w){\mathop{\rm sgn}} (x)(1 - \frac{8}{{\pi ^2 }}\sum\limits_{n = odd}^\infty  {\frac{{e^{ - \pi n\left| x \right|/w} }}{{n^2 }})} .
\end{eqnarray}

More generally, when the domain wall is located at $x = x_{\rm{DW}}$,

\begin{eqnarray}
V_H (x) &=& (\rho _H j_0 w){\mathop{\rm sgn}} (x-x_{\rm{DW}})(1 - \frac{8}{{\pi ^2 }}\sum\limits_{n = odd}^\infty  {\frac{{e^{ - \pi n\left| x-x_{\rm{DW}} \right|/w} }}{{n^2 }})}.
\end{eqnarray}

We now use eq. 24 to calculate \rxy$(H)$ in the presence of an in-plane external magnetic field. Although $H$ is nominally in the $xy$ plane during the experiment, in practice, there is always a slight misalignment towards $\hat{z}$ characterized by an angle $\delta \lesssim 1^{\circ}$ between $\vec{H}$ and $\vec{j}$.\cite{xiang_prb_2005} We divide our discussion of \rxy$(H)$ into four different regimes (see Fig. 2):
\begin{enumerate}
\item In regime I, the sample is in a single domain state while we sweep the external magnetic field from -2 T to $H_I = -5400$ Oe. The in-plane field is strong enough that the magnetization is completely aligned in the $xy$ plane and hence $R_{xy}=0$.
\item In regime II, the sample is still in a single domain state. As the in-plane magnetic field is further reduced ($\left|H \right|<\left|H_I\right|$) the magnetization of the sample starts to rotate toward the perpendicular direction, with the symmetry being broken by the slight misalignment. We assume the magnetization rotates coherently as a sine function of the external field, until the magnetization is totally aligned along $+\hat{z}$. During this process, the angle $\alpha$ between the magnetization and the $xy$ plane is:
\begin{eqnarray}
\alpha &=&  \frac{\pi }{2}(1-\frac{\left|H \right|}{H_I}),
\end{eqnarray}
and the Hall resistance measured at $x=0$ is given by:
\begin{eqnarray}
R_{xy} (H) &=& R_{H0} \sin \alpha,	
\end{eqnarray}

where $R_{H0}$ is the Hall resistance at zero field.

\item In region III, the external field changes sign and is swept from 0 to $H_I$. The $z$ component of the external field is now opposite to the magnetization of the sample, initiating magnetization reversal through the nucleation and propagation of DWs. For simplicity, we assume that a single DW starts from one end of the device and moves to the other end. Given the length of the actual Hall bar $L = 1500~ \mu$m, the domain wall is located at $x_{\rm{DW}}= \frac{L}{2}$. From the experimental data, the field at which \rxy $(H_C)=0$ marks the point at which $x_{\rm{DW}}=0$. We assume that the position of the DW varies linearly with $H$, \cite{tang_prb_2004, cheng_prl_2005} so that $x_{\rm{DW}} = \frac{L}{2}\times\frac{H-H_C}{H_C}$.	
Further, as shown above in eqs. (25) and (26), $R_{xy} (H)$ also changes because the out-of-plane magnetization rotates with the external in-plane field.

Then, the Hall resistance measured by the Hall probe at $x=0$ in region III is
\begin{eqnarray}
R_{xy} (H) &=& R_{H0}\sin{\frac{\pi }{2}(1-\frac{\left|H \right|}{H_I}) }{\mathop{\rm sgn}} (-x_{\rm{DW}})(1 - \frac{8}{{\pi ^2 }}\sum\limits_{n = odd}^\infty  {\frac{{e^{ - \pi n\left| x_{\rm{DW}} \right|/w} }}{{n^2 }})}.
\end{eqnarray}

\item In region IV, the sample is in a single-domain state again. The external field ($> H_I$) forces the magnetization to be fully aligned in the $xy$ plane and the Hall resistance stays at zero.

\end{enumerate}

We show a representative fit in figure 2. An identical calculation can be applied to the process when the magnetic field sweeps from positive to negative (not shown). The calculated transverse MR is in good agreement with the experimental results, indicating that the model used here is appropriate. 

\section {Simulation of the Longitudinal Magnetoresistance}

We now calculate \rxx$(H)$, noting that the value of \rxx~ depends on the relative locations of the domain wall and the electrodes. For two electrodes placed at points $x=l/2$ ($l=450~ \mu$m) along the lower edge ($y=0$), there exist three possible cases of $x_{\rm{DW}}$. To first order in $\beta$, \rxx~ is given by:

\begin{enumerate}
\item $x_{\rm{DW}}<-l/2$, 
\begin{eqnarray}
R_{xx}  &=& \frac{{V_2 ( - \frac{l}{2},0) - V_2 (\frac{l}{2},0)}}{{j_0 wt}} \nonumber\\
&=& R_S  + R_S (\sum)_{I}\\
(\sum)_{I}&=&\frac{{\rho _H }}{\rho }\frac{w}{l}\frac{4}{{\pi ^2 }}\sum\limits_{n = 1}^\infty  {\frac{{\exp ( - \frac{{n\pi }}{w}\left| {\frac{l}{2} + x_{\rm{DW}} } \right|) - \exp ( - \frac{{n\pi }}{w}\left| {\frac{l}{2} - x_{\rm{DW}} } \right|)}}{{n^2 }}}
\end{eqnarray}
where $R_S  = \rho l/(wt)$, $t$ is the thickness of the sample, and $w$ is the width of the Hall bar.

\item $-l/2<x_{\rm{DW}}<l/2$
\begin{eqnarray}
R_{xx}  &=& \frac{{V_1 ( - \frac{l}{2},0) - V_2 (\frac{l}{2},0)}}{{j_0 wt}} \nonumber\\ 
&=& R_S  - R_S [(\sum)_{II}-1]\\
(\sum)_{II}&=&\frac{{\rho _H }}{\rho }\frac{w}{l}\frac{4}{{\pi ^2 }}\sum\limits_{n = odd}^\infty  {\frac{{\exp ( - \frac{{n\pi }}{w}\left| {\frac{l}{2} + x_{\rm{DW}} } \right|) + \exp ( - \frac{{n\pi }}{w}\left| {\frac{l}{2} - x_{\rm{DW}} } \right|)}}{{n^2 }}}
\end{eqnarray}

\item $x_{\rm{DW}}>l/2$
\begin{eqnarray}
R_{xx}  &=& \frac{{V_1 ( - \frac{l}{2},0) - V_1 (\frac{l}{2},0)}}{{j_0 wt}} \nonumber\\ 
&=& R_S  - R_S (\sum)_{III} \\
(\sum)_{III}&=&\frac{{\rho _H }}{\rho }\frac{w}{l}\frac{4}{{\pi ^2 }}\sum\limits_{n = odd}^\infty  {\frac{{\exp ( - \frac{{n\pi }}{w}\left| {\frac{l}{2} - x_{\rm{DW}} } \right|) - \exp ( - \frac{{n\pi }}{w}\left| {\frac{l}{2} + x_{\rm{DW}} } \right|)}}{{n^2 }}}
\end{eqnarray}
Note that 
\begin{eqnarray}
\frac{{\rho _H}}{\rho} \frac{w}{l} &=& \frac{{\rho _H}}{t} \frac{wt}{\rho l} \nonumber \\
&=& {R_{xy} (H)}/R_{S}	
\end{eqnarray}

\end{enumerate}

Again, recall that $R_{xy} (H)$ varies with the out-of-plane rotation of the magnetization, as discussed in the last section.
\begin{eqnarray}
R_{xy} (H)  &=& R_{H0}\sin{\frac{\pi }{2}(1-\frac{\left|H \right|}{H_I})},
\end{eqnarray}

Overall, $\Delta R_{xx} = (R_{xx} - R_S)/(R_S)$ is given by
\begin{eqnarray}
\Delta R&=&\frac{{R_{H0} }}{R_S }\sin{\frac{\pi }{2}(1-\frac{\left|H \right|}{H_I})}[\frac{4}{{\pi ^2 }}(\sum)  - \theta (\frac{l}{2} - \left| {x_{\rm{DW}} } \right|)] \nonumber \\
\sum &=&\sum\limits_{n = odd}^\infty  {\frac{{\exp ( - \frac{{n\pi }}{w}\left| {\frac{l}{2} + x_{\rm{DW}} } \right|)-{\mathop{\rm sgn}} ({\left| x_{\rm{DW}}\right|  - \frac{l}{2}} )\exp ( - \frac{{n\pi }}{w}\left| {\frac{l}{2} - x_{\rm{DW}} } \right|)}}{{n^2 }}} ,	
\end{eqnarray}
where $x_{\rm{DW}} = \frac{L}{2}\times\frac{H-H_C}{H_C}$.

The calculated $\Delta R_{xx}$ is shown as the solid line in the figure 3 and is qualitatively in agreement with the experimental results. We speculate that the magnitude of the calculated anomalous spike is higher than that of the measured one because the measured magnetoresistance is sensitive to the actual structure of the DW. In our highly idealized model, the DW is assumed to be a perfect plane in the $yz$ plane, perpendicular to the long edge of the Hall bar. In reality, the DW structure is likely to be far more complicated, as suggested by recent magneto-optical imaging of the easy axis magnetization reversal process of tensile-strained \gma~ samples.\cite{douriat} The reasonable agreement between the model and experiment is hence quite surprising and better than might be anticipated. 

Finally, it is important to compare these results with the antisymmetric anomalies reported in the metallic multilayer samples with perpendicular magnetic anisotropy. In that case, the antisymmetric magnetoresistance s observed for magnetic fields applied along the easy axis (perpendicular to the sample plane). However, in the \gma~ samples studied here, we do not observe any such antisymmetric magnetoresistance in the perpendicular geometry.\cite{xiang_prb_2005} Noting that the coercive field for easy axis magnetization reversal in tensile-strained \gma~ is very small ($\sim$ 20 Oe), we speculate that DW nucleation and propagation occurs very rapidly during easy axis magnetization reversal. The absence of the MR anomalies in the perpendicular geometry can then be attributed to the lack of experimental resolution in current experiments. In contrast, in the planar geometry, the effective coercive field for magnetization reversal is much larger ($\sim$ 2200 Oe) so that magnetization reversal occurs adiabatically with a very slow nucleation and propagation of DWs across the sample. Finally, we note again that magneto-optical images of the easy axis magnetization reversal process show a very complicated domain nucleation and propagation that probably statistically averages out the contributions to \rxx~ from circulating currents.  

\section {Summary}

In summary, we have shown that -- in the planar geometry -- the key features of the unusual longitudinal and transverse magnetoresistance in tensile-strained \gma~ can be readily explained by extending an earlier model applied to easy axis magnetoresistance in ferromagnets. Since the current model addresses the hard axis magnetoresistance in samples with perpendicular anisotropy, the interplay between the magnetization reversal process and magnetoresistance is more complicated than in earlier studies. Our model reveals that there are two different magnetic states involved in the sample during the magnetic field sweep: one is a single-domain state, when the in-plane magnetic field is strong enough to align the magnetization in the $xy$ plane, or when the magnitude of the magnetic field decreases from maximum to zero and the magnetization of the sample spontaneously rotates from the $xy$ plane to the $z$ axis. The other state is a two-domain state, ideally separated by a single DW. When the magnetic field changes sign, the $z$ component of the in-plane field, due to the unintentional misalignment, initiates DW nucleation and propagation in the sample. The observed field-antisymmetric anomalies and unusual Hall loops then arise from AHE contributions when a DW is located in between the voltage probes.

This research has been supported by grant numbers ONR N0014-05-1-0107 and NSF DMR-0305238. 
\begin{center}
{\bf References}
\end{center}

\newpage

\begin{center}
{\bf Figure Captions}
\end{center}

Fig. 1. Schematic of a thin film with one $180^o$ domain wall at x $=$ 0. The film is assumed to be infinitely long. The width and the thickness of the film are $w$ and $t$, respectively.

Fig. 2. Measured (solid line) and calculated (open circles) transverse magnetoresistance in a tensile-strained \gma~ epilayer at $T = 80 $K. The cartoons show the magnetization configuration with respect to the Hall bar at different fields. The circle locates the field at which the Hall voltage is zero, indicating that the DW is located between midway between a pair of voltage probes.

Fig. 3. Measured (solid line) and calculated (open circles) longitudinal magnetoresistance in a tensile-strained \gma~ epilayer at $T = 80 $K. The cartoons show the magnetization configuration with respect to the Hall bar at different fields. 

\newpage
\begin{figure}[h]
\begin{center}
\includegraphics[]{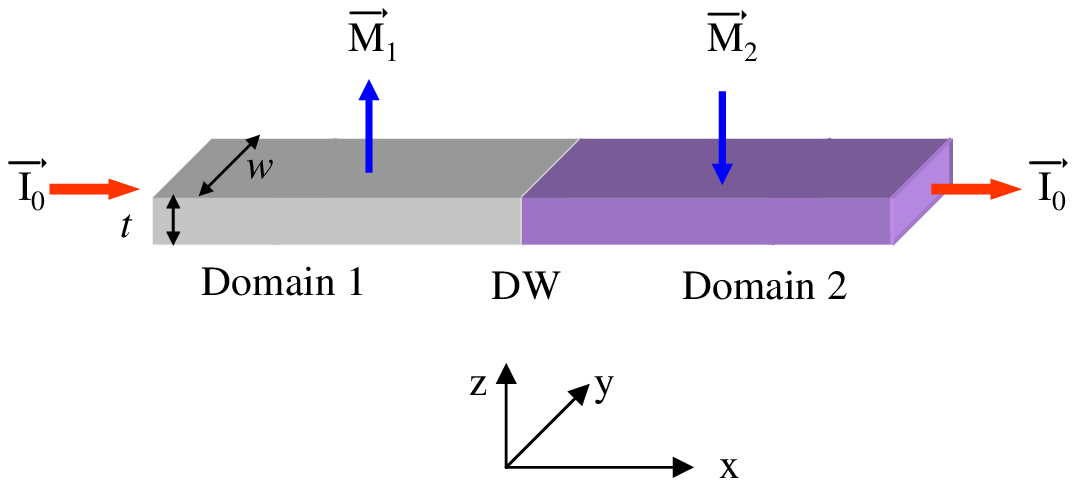}
\label{gang_fig1}
\end{center}
\end{figure}

\newpage
\begin{figure}[h]
\begin{center}
\includegraphics[]{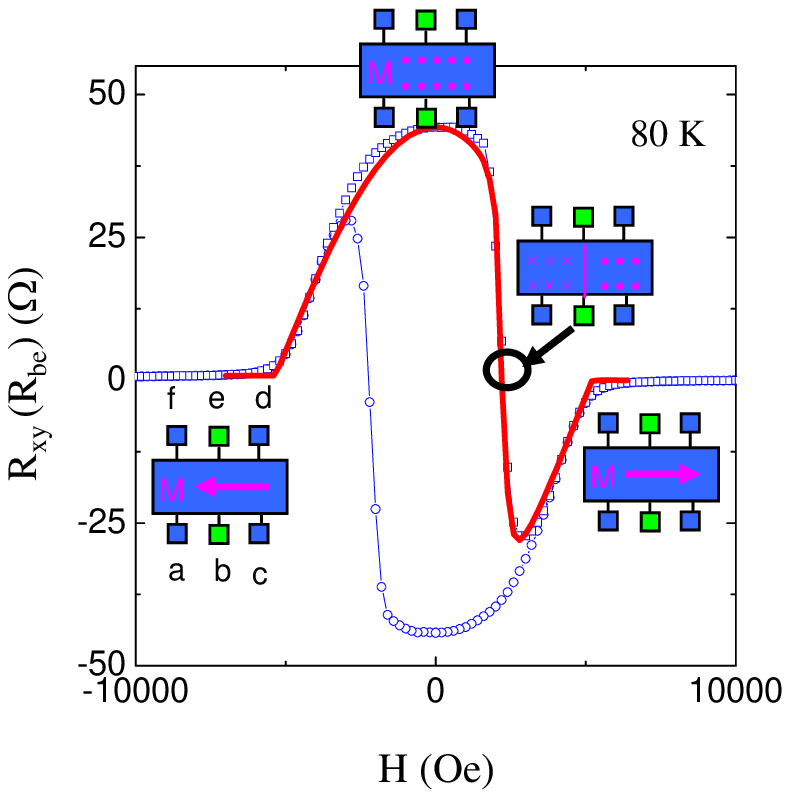}
\label{gang_fig2}
\end{center}
\end{figure}

\newpage
\begin{figure}[h]
\begin{center}
\includegraphics[]{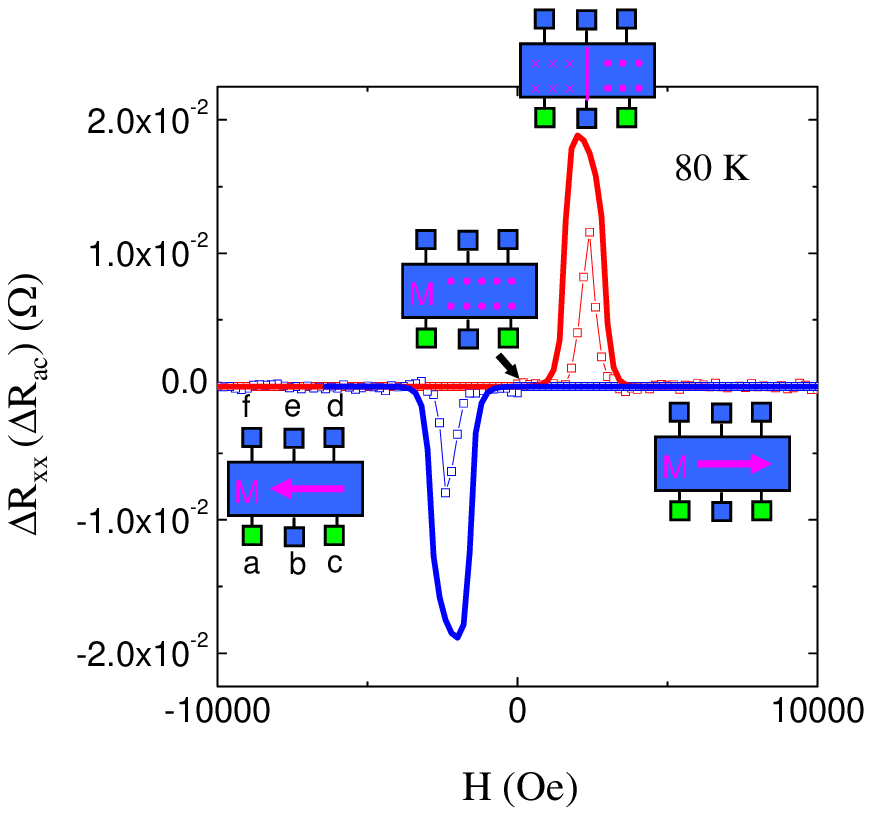}
\label{gang_fig3}
\end{center}
\end{figure}

\end{document}